\documentclass{LMCS}

\def\doi{8 (2:08) 2012}
\lmcsheading%
{\doi}
{1--15}
{}
{}
{Jan.~16, 2012}
{Jun.~\phantom01, 2012}
{}

\usepackage{enumerate}
\usepackage{hyperref}
\usepackage{amssymb,verbatim}
\usepackage{array,amsmath}
\def\restrict{\mathpunct\restriction}

\usepackage{graphicx}

\newcommand{\turnstyle}{\vdash}

\def\restrict{\mathpunct\restriction}

\begin{document}
\title[Width and size of regular resolution proofs]{Width and size of regular resolution proofs}
\author[A.~Urquhart]{Alasdair Urquhart}
\address{University of Toronto}
\email{urquhart@cs.toronto.edu}
\thanks{The author gratefully acknowledges
the support of
the Natural Sciences and Engineering Research Council of Canada.}

\keywords{regular resolution proofs, size of proofs, width of proofs}
\subjclass{F2.2,F4.1}

%%%%%%%%%%%%%%%%%%%%%%%%%%%%%%%%%%%%%%%%%%%%%%%%%%%%%%%%%%%%%%%%%%%%%%%%%%%

\begin{abstract}
This paper discusses the topic of the minimum width of a regular resolution refutation
of a set of clauses.
The main result shows that there are examples having small regular 
resolution refutations, for which any regular refutation must contain
a large clause.
This forms a contrast with corresponding results for general resolution refutations.
\end{abstract}

\maketitle

\section*{Introduction}
Recent results \cite{alekjohpiturq2007,urquhart2011}
showing near-exponential separations between the size 
of regular and general refutations of certain sets of clauses
also show a separation of general and regular
resolution width.
That is to say, the examples used in showing the size separation
have large regular resolution width, but bounded general
resolution width.  

This observation suggests that it might be possible to 
prove results for regular resolution similar to those of 
Ben-Sasson and Wigderson \cite{bens01} for tree resolution
and general resolution.  
The main theorem below shows that this hope is bound to be
disappointed; it exhibits examples
having small regular resolution size, but large regular width.

The first part of the paper gives a characterization of regular resolution width, in the 
style of Atserias and Dalmau \cite{atseriasdalmau2008}.
The second part discusses the relationship between the size and width of
regular resolution refutations.

\section{Resolution proofs and their width}

A \emph{literal} is a propositional variable $x$ 
or its negation $\lnot x$.
A \emph{clause} is a set of literals, interpreted as the disjunction of the set.
For clauses containing exactly one positive literal,  we use
the implication $p_1, \dots, p_k \rightarrow q$ as  alternative notation
for the clause $\neg p_1 \vee \dots \vee \neg p_k \vee q$.
For notational convenience, we shall also allow the case where the positive
literal $q$ is replaced by the propositional constant $\bot$.
For any assignment $\sigma$, $\sigma(\bot) = 0$, so that the expression
``$p_1, \dots, p_k \rightarrow \bot$'' is an alternative notation for
the purely negative clause $\neg p_1 \vee \dots \vee \neg p_k$ 

If $p$ is a variable, and $C$ a clause, then we say that $p$ has a 
{\em positive occurrence} in $C$ if $p$ is in $C$, and a 
{\em negative occurrence} in $C$ if $\neg p$ is in $C$.
In addition, we shall say that $\bot$ has a positive occurrence in 
the purely negative Horn clause $p_1, \dots, p_k \rightarrow \bot$.
If $\Sigma$ is a set of clauses, and $x,y$ are variables in $\Sigma$, 
or the propositional constant $\bot$,  then we
say that there is an {\em implicational chain from $x$ to $y$ in $\Sigma$} 
if there is a sequence $x = x_0, \dots, x_k = y$ of variables (or constants) and a sequence
$C_1, \dots, C_k$ of clauses so that for all $i$, $0 < i  \leq k $,
$x_{i-1}$ occurs negatively and $x_i$  positively in $C_i$.

The \emph{resolution rule} allows us to derive 
the \emph{resolvent} $C \vee D$ from the clauses $C \lor x$ and $D \lor \lnot x$ 
by \emph{resolving on} the variable $x$.
 A \emph{resolution derivation} of a clause $C$ from a set of clauses $\Sigma$  consists of a
sequence of clauses in which each clause is either a clause of $\Sigma$, or
derived from earlier clauses by resolution,  and $C$ is the last clause in the
sequence; it is a \emph{refutation} of $\Sigma$ if $C$ is the empty clause
$\Lambda$.  

The {\em size} of a resolution proof is the number of occurrences of clauses in the proof --
that is to say, the length of the proof considered as a sequence.
For a contradictory set of clauses $\Sigma$, we write $S(\Sigma)$ for
the minimum size of a resolution refutation of $\Sigma$.
A resolution proof is a {\em tree-style} proof if every clause
in the proof is used at most once as a premiss in a resolution inference.
We write $S_T(\Sigma)$ for the minimum size of a tree-style refutation
of a contradictory set of clauses $\Sigma$.
A resolution refutation of a set of clauses $\Sigma$ is an {\em input proof} if 
in every application of the resolution rule in it, at least one 
premiss of the application is an input clause in $\Sigma$.
Every input refutation is automatically a tree-style refutation.

An {\em irregularity} in a resolution proof is a sequence of clauses
$C_1, \dots, C_k$ so that $C_{i+1}$ is derived from $C_i$ (that is,
$C_i$ is one of the premisses of a resolution inference in which
the conclusion is $C_{i+1}$), and there is a variable 
that occurs in $C_1$ and $C_k$, but not in any intermediate clause
$C_j$, $1 < j < k$. 
A resolution proof is {\em regular} if it contains no irregularity.

If $V$ is a set of propositional variables, then an {\em assignment}
is a Boolean function defined on a subset of $V$, that is, an 
assignment of $\{ 0,1 \}$ to some or all of the variables in $V$.
If $\alpha$ is an assignment, then we write $| \alpha |$ for the
cardinality of $\alpha$, the number of variables to which
$\alpha$ assigns values.
The result of restricting a clause $C$ by setting a literal $l$ is defined
as follows.
If the literal $l$ occurs in $C$, then $C[l := 1] = 1$, while $C[l := 0]$ is
$C \setminus \{ l \}$.
If $\Sigma$ is a set of clauses, and $v \in \{ 0,1 \}$, then $\Sigma[l := v]$ is the set of
clauses $\{ C[l := v] : C \in \Sigma \} \setminus \{ 1 \}$.

The {\em width} of a clause is the number of literals in it.
The width $w(\Sigma)$ of a set of clauses is the maximum width of a clause
in $\Sigma$, while 
the width of a resolution proof is the maximum width of a clause in it.
If $\Sigma$ is a contradictory set of clauses, then we define the 
{\em refutation width} of $\Sigma$, written $w(\Sigma \vdash 0)$, to be the
minimum width of a resolution refutation of $\Sigma$.
If $\mathcal F$ is a family of resolution proofs, we define a restricted notion
of refutation width, the {\em $\mathcal F$-refutation width}, $w(\Sigma \vdash^{\mathcal F} 0)$, to 
be the minimum width of a refutation of $\Sigma$ that belongs to $\mathcal F$.
In particular, we define the {\em regular refutation width}, $w(\Sigma \vdash^{\mathcal R} 0)$,
of a contradictory
set of clauses $\Sigma$ to be the $\mathcal R$-refutation width where $\mathcal R$ is the class of
all regular resolution proofs.

The notation $\log x$ stands for the base two logarithm of $x$,
and $\log^k x$ for $(\log x)^k$.

\section{Characterization of general resolution width}

In this section, we give a proof of a result of Atserias and Dalmau
\cite{atseriasdalmau2008} characterizing the width of general resolution refutations.
The characterization is in terms of a two player game, that we shall
call the {\em $k$-width game}, played by the {\em Prover} and the {\em Adversary}.
\footnote{Atserias and Dalmau, following the tradition of finite model theory,
call their players the {\em Spoiler} and the {\em Duplicator}, but our 
terminology seems clearer in the present context.}
The rules of the game are as follows.

The players are given a set of clauses $\Sigma$, on 
a set $V$ of variables, and an integer parameter $k \geq 0$.
The players together construct a succession of assignments to
the variables in $V$.
Initially, the assignment is empty.
Each round of the game proceeds as follows, starting from
a current assignment.
First, the Prover queries an unassigned variable, and 
the Adversary assigns a value to it.
Second, the Prover is allowed to delete some of the values of the 
variables in $V$ from the assignment resulting from the Adversary's reply;
the result is the new current assignment.

The Adversary can win in two ways.
First, if the current assignment (after deletions) assigns values 
to more than $k$ variables;
second, if an earlier assignment is repeated during the
play of the game.
The Prover wins if the current assignment falsifies an initial clause in $\Sigma$.
Clearly every play of the game must eventually terminate with a win for the
Prover or for the Adversary (Atserias and Dalmau define their game so that when
the Adversary wins, the game can continue infinitely).

\begin{defi}
\label{extendiblefamilydefinition}
If $\Sigma$ is a set of clauses on a set $V$ of variables, then a non-empty
family $\mathcal A$ of $V$-assignments is an {\em extendible $k$-family for $\Sigma$} if it
satisfies the following conditions:
\begin{enumerate}[(1)]
\item
No assignment in $\mathcal A$ falsifies a clause in $\Sigma$;
\item
Each assignment $\alpha$ in $\mathcal A$ satisfies the condition $| \alpha | \leq k$;
\item
If $\alpha \in {\mathcal A}$, and $\beta \subseteq \alpha$, then $\beta \in {\mathcal A}$;
\item
If $\alpha \in {\mathcal A}$, $| \alpha | < k$, and $x \in V$, then there is a $\beta \in {\mathcal A}$,
so that $\alpha \subseteq \beta$, and $\beta(x)$ is defined.
\end{enumerate}
\end{defi}

The next theorem shows that a resolution refutation of width $k$ constitutes
a winning strategy for the Prover, while an extendible $k+1$-family 
provides a winning strategy for the Adversary.

\begin{thm}
\label{widthcharacterizationtheorem}
{\bf [Atserias and Dalmau 2003]} 
Let $\Sigma$ be a set of clauses, and $k \geq w(\Sigma)$.
Then the following are equivalent:
\begin{enumerate}[\em(1)]
\item
There is no resolution refutation of $\Sigma$ of width $k$;
\item
There is an extendible $k+1$-family for $\Sigma$;
\item
The Adversary wins the $k+1$-width game based on $\Sigma$.
\end{enumerate}
\end{thm}

\proof
First, let us suppose that there is no resolution refutation of $\Sigma$
of width $k$.
Let $\mathcal C$ be the set of all clauses having a resolution proof from $\Sigma$
of width at most $k$; 
since $w(\Sigma) \leq k$,  $\Sigma \subseteq {\mathcal C}$.
Let $\mathcal A$ be the set of all assignments of size at most $k+1$ that do not
falsify any clause in $\mathcal C$.
We claim that $\mathcal A$ is an extendible $k+1$-family for $\Sigma$.
First, $\mathcal A$ is non-empty, because it contains the empty assignment
(since $\mathcal C$ does not contain the empty clause).
Second, $\mathcal A$ satisfies the first three conditions
of Definition \ref{extendiblefamilydefinition}, by construction.
To prove the fourth condition, let $\alpha \in {\mathcal A}$, and 
$| \alpha | \leq k$, $x \in V$, but there is no extension $\beta$ of $\alpha$
in $\mathcal A$ with $\beta(x)$ defined.
It follows that there is a clause $D \in {\mathcal C}$ that 
is falsified if we extend $\alpha$ by setting $x$ to $0$.
Then $D = E \vee x$ for some $E$, since otherwise $\alpha$ would 
falsify $D$.
Similarly, there is a clause $F \vee \neg x$ in $\mathcal C$
that is falsified by the extension of $\alpha$ that sets  $x$ to 1.
But then $\alpha$ must falsify $E \vee F$, showing that $E \vee F$ has 
width at most $k$, since $|\alpha| \leq k$.
Hence, it follows that $E \vee F$ is in
$\mathcal C$, contradicting our assumption that $\alpha$ is in $\mathcal A$.

Second, let us suppose that there is an extendible $k+1$-family for $\Sigma$.
Then the Adversary can play the $k$-width game on $\Sigma$ by responding
to the Prover's queries with the appropriate assignment from the family,
starting with the empty assignment.
Since no assignment in the family falsifies an initial clause, this
strategy must eventually end in a win for the Adversary, no matter how
the Prover plays.

Finally, let us suppose that there is a resolution refutation of $\Sigma$ of
width $k$.
Then the refutation provides the Prover with a winning strategy
in the $k+1$-width game based on $\Sigma$.
Starting from the empty clause at the root, the Prover follows a path
in the refutation so that at each round, the assignment (after appropriate deletions) 
is a minimal assignment falsifying the current clause.
The variable queried is the variable resolved upon to derive the current
clause.
This strategy must result in a win for the Prover
when the path reaches a clause in $\Sigma$.
\qed

 \section{Characterization of regular resolution width}

In the present section, we modify the result of Atserias and Dalmau
to characterize the width of regular resolution refutations.
The characterization is again in terms of a two player game, that we shall
call the {\em regular $k$-width game}.
The game is exactly the same as that described in the previous
section, but with the added condition that the Prover can never
query a previously queried variable.

As in the case of general resolution width, we can characterize the regular
resolution width in terms of extendible families of assignments.
However, we need to redefine the notion of an assignment.
In the earlier notion of assignment, a variable could be in three states:
{\em positive} (1), {\em negative} (0), and {\em unassigned} ($\ast$).
For the case of regular resolution, we define an {\em extended assignment}
to be an assignment of values in which each variable can be in four states:
{\em positive} (1), {\em negative} (0), {\em unassigned} ($\ast$), 
or {\em forgotten} ($\boxtimes$).
The {\em empty extended assignment} to a set $V$ of variables 
consists of the assignment in which all variables in $V$ are unassigned ($\ast$)
(this should be distinguished from assignments in which all of
the variables are unassigned or forgotten ($\boxtimes$)).

If $\alpha$ is an extended assignment, then those variables that are assigned
the values 0 or 1 are the {\em live} variables in $\alpha$, and we write
$| \alpha |$ for the number of live variables in $\alpha$.
If $\alpha$ and $\beta$ are extended assignments to a set of variables $V$, then
we write $\alpha \subseteq \beta$ if $\beta$ results from $\alpha$ by 
replacing some unassigned variables by live variables.
We also write $\alpha \sqsubseteq \beta$ if $\alpha$ results from $\beta$
by forgetting some variables, that is, changing the value of a live
(0 or 1) variable to $\boxtimes$.

As in the case of the earlier $k$-width game, 
the players are given a set of clauses $\Sigma$, on 
a set $V$ of variables, and an integer parameter $k \geq 0$.
Together, they construct a succession of extended assignments to
the variables in $V$.
Initially, the assignment is empty.
Each round of the game proceeds as follows.
First, the Prover queries an unassigned variable, and 
the Adversary assigns a value to it.
Next, the Prover is allowed to forget some of the variables in
the resulting assignment, that is, to
change the value of a live variable
from 0 or 1 to $\boxtimes$ (forgotten);
the result is the new current assignment.

Again, the Adversary can win in two ways.
First, if the current assignment assigns values 
to more than $k$ variables;
second, if the Prover has not won up to this point, but there are no unqueried
variables, so the Prover has no legal move.
The Prover wins if the current assignment falsifies an initial clause in $\Sigma$
(if this clause contains more than $k$ variables, then we count this as a win for the Adversary).
As before, every play of the game must eventually terminate with a win for the
Prover or for the Adversary.

\begin{defi}
\label{regularextendiblefamilydefinition}
If $\Sigma$ is a set of clauses on a set $V$ of variables, then a 
family $\mathcal A$ of extended $V$-assignments is a
{\em regular extendible $k$-family for $\Sigma$} if it
satisfies the following conditions:
\begin{enumerate}[(1)]
\item
The empty assignment belongs to $\mathcal A$;
\item
No assignment in $\mathcal A$ falsifies a clause in $\Sigma$;
\item
Each assignment $\alpha$ in $\mathcal A$ satisfies the condition $| \alpha | \leq k$;
\item
If $\alpha \in {\mathcal A}$, and $\beta \sqsubseteq \alpha$, then $\beta \in {\mathcal A}$;
\item
If $\alpha \in {\mathcal A}$, $| \alpha | < k$, $x \in V$, and $\alpha(x) = \ast$, 
then there is a $\beta \in {\mathcal A}$,
so that $\alpha \subseteq \beta$, and $\beta(x) = 0$ or $\beta(x) = 1$.
\end{enumerate}
\end{defi}

\noindent
The next theorem is the analogue of Theorem \ref{widthcharacterizationtheorem}
for regular resolution.

\begin{thm}
\label{regularwidthcharacterizationtheorem}
Let $\Sigma$ be a set of clauses, and $k \geq w(\Sigma)$.
Then the following are equivalent:
\begin{enumerate}[\em(1)]
\item
There is no regular resolution refutation of $\Sigma$ of width $k$;
\item
There is a regular extendible $k+1$-family for $\Sigma$;
\item
The Adversary wins the regular $k+1$-width game based on $\Sigma$.
\end{enumerate}
\end{thm}

\noindent
\proof
( 1 $\Rightarrow$ 2 ):
Let us suppose that there is no regular resolution refutation of $\Sigma$
of width $k$.
Define $\mathcal C$ to be the set of all clauses having a regular resolution proof from $\Sigma$
of width at most $k$; 
since $w(\Sigma) \leq k$,  $\Sigma \subseteq {\mathcal C}$.
Let $\mathcal A$ be the set of all extended assignments of size at most $k+1$ that do not
falsify any clause in $\mathcal C$.
We claim that $\mathcal A$ is an extendible $k+1$-family for $\Sigma$.

Since the empty clause $\Lambda$ does not belong to $\mathcal C$,
the empty assignment is in $\mathcal A$, so the first condition in
Definition \ref{regularextendiblefamilydefinition} 
is satisfied.
The second condition holds because $\Sigma \subseteq {\mathcal C}$, 
and the third condition by definition.
The fourth condition also follows from the definition of $\mathcal A$.

It remains to prove the fifth condition.
Assume that $\alpha \in {\mathcal A}$, and 
$| \alpha | \leq k$, $x \in V$, and $\alpha(x) = \ast$,  but there is no extension $\beta \supseteq \alpha$
in $\mathcal A$ with $\beta(x)$ defined.
Let $\alpha^0$ and $\alpha^1$ be the extended assignments obtained from
$\alpha$ by setting $x$ to 0 and 1, respectively.
Since neither $\alpha^0$ nor $\alpha^1$ belong to $\mathcal A$,
it follows that there are regular resolution derivations ${\mathcal R}^0$ and ${\mathcal R}^1$ of
clauses $C_0$ and $C_1$, each having width at most $k$, so that
for $i = 0,1$, $\alpha^i(C_i) = 0$.
Since $\alpha \in {\mathcal A}$, it follows that $C_0 = D \vee x$, and
$C_1 = E \vee \neg x$, for some clauses $D$ and $E$.
However, if we extend the regular resolution derivations ${\mathcal R}^0$ and ${\mathcal R}^1$ 
by resolving on $x$, so that the final clause is $D \vee E$, the result is a regular
resolution derivation of $D \vee E$, where $\alpha(D \vee E) = 0$.
Since $|\alpha| \leq k$, $D \vee E$ has at width at most $k$, showing that
$D \vee E$ is in $\mathcal C$; this contradicts our assumption that $\alpha \in {\mathcal A}$.

( 2 $\Rightarrow$ 3 ):  Second, let us suppose that there is a regular extendible $k+1$-family for $\Sigma$.
Then the Adversary can play the $k$-width game on $\Sigma$ by responding
to the Prover's queries with the appropriate assignment from the family,
starting with the empty assignment.
Since no assignment in the family falsifies an initial clause, this
strategy must eventually end in a win for the Adversary, no matter how
the Prover plays.

( 3 $\Rightarrow$ 1 ):
Finally, let us suppose that there is a regular resolution refutation of $\Sigma$ of
width $k$.
Then the refutation provides the Prover with a winning strategy
in the regular $k+1$-width game based on $\Sigma$.
Starting from the empty clause at the root, the Prover follows a path
in the refutation so that at the end of each round, after the Prover has
forgotten certain live variables, the remaining live variables are the domain of
a minimal assignment falsifying the current clause.
The variable queried is the variable resolved upon to derive the current
clause.
This strategy must result in a win for the Prover.
 \qed

\begin{cor}
The question ``Is there a regular resolution refutation of the set of 
clauses $\Sigma$ with width $k$?" is in PSPACE.
\end{cor}

\noindent
\proof
Theorem \ref{regularwidthcharacterizationtheorem}
shows that this question can be answered by an alternating Turing machine
operating in polynomial time. \qed

\medskip

In the case of general resolution width,  it is not clear whether the
corresponding problem is in PSPACE, because there is no polynomial
upper bound on how long the $k$-width game might last.

\section{Size and width of regular resolution proofs}
\subsection{The width and size of resolution proofs}

Recent results on size separation between regular and general 
resolution also show a width separation.

\begin{thm}
For each $n > 0$, 
there is a contradictory set of clauses with $O(n^2)$ variables
and $O(n^3)$ clauses for which the general resolution width
is bounded, but the regular resolution width is $\Omega(n)$.
\end{thm}

\noindent
\proof
The paper \cite{alekjohpiturq2007} 
implicitly contains such a separation.
More specifically the family of clauses $GT_{n,\rho}'$ defined in \S 3 of 
\cite{alekjohpiturq2007} fulfil the conditions of the theorem.
The $\Omega(n)$ lower bound on regular resolution width is proved
(implicitly) in Theorem 3.10 of that paper, which shows
an exponential ($2^{n/200}$) lower bound on the size of regular
resolution refutations of $GT_{n,\rho}'$.
\qed

\medskip

The author's paper \cite{urquhart2011} demonstrates an improved
size separation between regular and general resolution; it also shows
a width separation between the two forms of proof system.
The main theorem shows that for infinitely many $n$,
there is a set $\Pi_n$ of $O( n \log^5 n)$ clauses containing $O( n \log \log n)$
variables, where the maximum width of a clause in $\Pi_n$ and
the general resolution width are both $O(\log \log n)$,
while the regular resolution width is $\Omega(n / \log n)$.

The results just described suggest a natural conjecture 
that a good lower bound on the regular
width of a set of clauses leads to a good lower bound on the
size of a regular refutation of them.
For both general and tree resolution, Ben-Sasson and Wigderson
\cite{bens01} have proved strong results along these lines.

\begin{thm}
\label{BenSassonWigdersonTheorem}
{\bf [Ben-Sasson and Wigderson 2001]}
Let $\Sigma$ be a contradictory set of clauses with an underlying
set of variables $V$.
Then:
\begin{enumerate}[\em(1)]
\item
$S_T(\Sigma) \geq 2^{ w(\Sigma \vdash 0 ) - w(\Sigma)}$;
\item
$S(\Sigma) = \exp \left( \Omega  \left(
 \frac{(w(\Sigma \turnstyle 0) - w(\Sigma))^{2}}
{|V|} \right) \right)$
\end{enumerate}
\end{thm}

\medskip

Given the width and size separation results between regular and general resolution
cited above, it seems reasonable to conjecture that the second lower bound
proved by Ben-Sasson and Wigderson might hold, in the form where we replace
``resolution size'' by ``regular resolution size,'' and 
``resolution width'' by ``regular resolution width.''
In the remainder of the paper, we show that this conjecture fails.

Before proceeding to the main constructions, it may be helpful to the reader
to clarify the relations between the various forms of resolution discussed here.

If we consider the size measure alone, then it is not hard to see that 
regular resolution is at least as powerful as tree resolution.
This is because a pruning procedure \cite{ts70} \cite[p.\ 436]{urq95} 
can be applied to a tree refutation to remove any irregularities 
while decreasing the size of the tree.
On the other hand, the lower bound on width for regular refutations 
proved below does not apply to tree resolution, since the minimum
width of a tree-style refutation of a set of clauses is the same
as that of a general resolution refutation (we can convert
any general resolution proof into a tree-style proof by 
repeating subderivations).

However, if we insist on restricting our attention to tree-style
refutation {\em of minimum size}, then the lower bounds
on width do apply, since such refutations are necessarily 
regular.

\section{Pebbling games and pebbling formulas}
\label{pebblingsection}
\subsection{The pebbling game}
A {\em pointed graph} $G$  is a directed acyclic graph 
where all vertices have indegree at most two, 
having a unique sink, or target vertex,
to which there is a directed path from all the vertices in $G$.
It is {\em binary} if all vertices except for
the source vertices have indegree two.
If $v$ is a vertex in a pointed graph $G$, then
$G \restrict v$ is the subgraph of $G$ restricted to the
vertices from which there is a directed path to $v$.

The {\em pebbling game} played on a pointed graph $G$ 
is a one-player game in the course of which pebbles are placed on or
removed from vertices in $G$.
The rules of the game are as follows;
\begin{enumerate}[(1)]
 \item 
A pebble may be placed on a source vertex  at any time.
\item
If all predecessors of a vertex are marked with pebbles, then a pebble may be placed
on the vertex itself.
\item
A pebble may be removed from a vertex at any time.
\end{enumerate}
A {\em move} in the game
consists of placing or removing one of the  pebbles
in accordance with one of the three rules.
The {\em configuration} at a given stage in the game is the set of vertices
in $G$ that are marked with a pebble. 
A play of the game begins with no pebbles on $G$.
The goal of the game is to place a pebble on the sink vertex $t$, while
minimizing the number of  pebbles used (that is, minimizing the number of pebbles
on the graph at any stage of the game).
Thus a successful play of the game can be presented as a sequence of configurations $C_0, \dots, C_k$,
where $C_0 = \emptyset$ and $t \in C_k$, where $C_{j+1}$ is obtained from
$C_j$ by one of the three rules.

A {\em strategy} for the game is a sequence of moves following the 
rules of the game that ends in pebbling the target vertex.
The {\em cost} of such a strategy is the minimum number of pebbles
required in order to execute it, that is to say, the size of the largest 
configuration in the sequence of configurations produced by following the strategy.
The {\em pebbling number} of $G$, written as $\sharp G$, 
is the minimum cost of a strategy for the pebbling game played on $G$.

\subsection{Pebbling formulas}
We associate a contradictory set of clauses $\mbox{Peb}(G)$ with every pointed graph $G$.
Each vertex in $G$ except the target $t$  is assigned a distinct variable;
to simplify notation, we identify a vertex with the variable associated with it, 
and use the notation $\mbox{Var}(G)$ for the set of these variables.
We associate the constant $\bot$ (falsum) with the target vertex $t$,
and make the identification $t = \bot$.

\begin{defi}
 If $G$ is a pointed graph,  $\mbox{Peb}(G)$ is a set of clauses
expressed in terms of the variables $\mbox{Var}(G)$, 
so that $\mbox{Peb}(G) = \{ \mbox{Clause}(v) : v \in G  \}$.
\begin{enumerate}[(1)]
 \item 
If $v$ is a source vertex of $G$,  then $\mbox{Clause}(v) = v$.
\item
If $v$ is a vertex in $G$, with predecessor $u$, then 
$\mbox{Clause}(v) = u\rightarrow v$.
\item
If $v$ is a vertex in $G$, with predecessors $u,w$, then 
$\mbox{Clause}(v) = u, w \rightarrow v$.
\end{enumerate}
\end{defi}

If we set some variables in $\mbox{Peb}(G)$, then the resulting set of clauses
is not necessarily of the form $\mbox{Peb}(G')$, where $G'$ is a subgraph
of $G$.
We shall focus on a family of special assignments, called
{\em pebbling assignments},  that preserve this property.
If $v \in G$, $v \neq t$, then we define the assignment $[\![ v : = 1 ] \! ]$
to be the assignment defined by first setting the variable $v$ to 1, and then
setting to 1 any variable $u$ for which there is no implicational chain
from $u$ to $\bot$ in the resulting clause set.
The assignment $[\![ v : = 0 ] \! ]$ is defined as follows:
first, choose a directed path $\pi = (v, \dots, t)$ from $v$ to the
target $t$, set all the vertices in the path to 0, and in addition
set any vertex from which $v$ is not reachable, but not in the path $\pi$,  to 1.
The assignment $[\![ v : = 0 ] \! ]$ is not uniquely determined by this construction, since
it depends on the path chosen -- however, this is not important, since
the set of clauses $\mbox{Peb}(G)  \restrict  [\![ v : = 0 ] \! ]$
resulting from the restriction is independent of the path.
A {\em pebbling assignment} results from a sequence of restrictions of the form 
$[\![ v : = 0 ] \! ]$ and $[\![ w : = 1 ] \! ]$.

The effect of the restrictions just defined can be described directly as an 
operation on the underlying graph.
If $G$ is a pointed graph, and $v \in G$, $v \neq t$, $G[ v := 1 ]$ is the 
graph resulting from $G$ by first removing $v$, together with all edges entering or 
leaving $v$, and then restricting the resulting graph to the vertices from which
the target vertex $t$ is accessible. 
$G[ v := 0 ]$ is the pointed graph $G \restrict v$.

\begin{lem}
\label{pebblinglemma} \hfill
\begin{enumerate}[\em(1)]
 \item 
For $b = 0, 1$, $\mbox{Peb}(G)  \restrict [\![ v : = b ] \! ] = \mbox{Peb}(G[ v := b ])$.
\item
 If $G$ is a pointed graph, and $v \in G$, then 
\[
 \sharp G \leq  \max \{ \sharp G[ v := 0 ],  \sharp G[ v := 1 ]  + 1 \}.
\]
\end{enumerate}
\end{lem}

\proof
The first part of the lemma follows straightforwardly from the definitions.
For the second part, 
we employ the following strategy in the pebble game on $G$;
the strategy is the same as the one used in Lemma 15 of \cite{bensassonimpagliazzowigderson2004}.

First, follow a minimum cost strategy to pebble $v$ in $G[ v := 0 ]$.
Second, leaving a pebble on $v$, but removing all other pebbles, follow a minimum
cost strategy in the pebbling game on $G[ v := 1 ]$ to pebble the 
target vertex in $G$, using the extra pebble for any moves where 
a pebble is needed on $v$ to justify a placement.
The cost of this strategy is at most $  \max \{ \sharp G[ v := 0 ],  \sharp G[ v := 1 ]  + 1 \}$.
\qed

\medskip

\begin{figure}
\begin{center}
\leavevmode
\includegraphics[height=7cm]{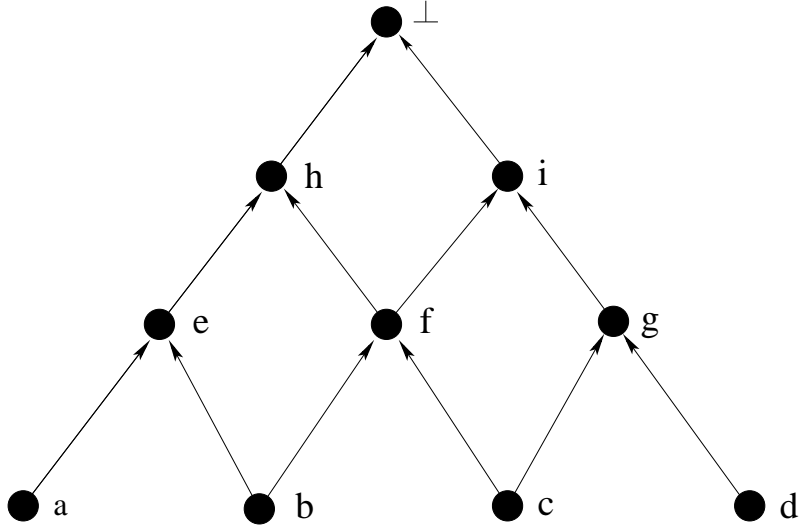}
\caption{A pyramid graph}
\end{center}
\end{figure}

\begin{exa}
If $G$ is the pyramid graph shown in Figure 1, then $\mbox{Peb}(G)$ is the set of clauses 
\[
\{ a,b,c,d,  (a,b \rightarrow e), (b,c \rightarrow f), (c,d \rightarrow g), (e,f \rightarrow h), (f,g \rightarrow i),
    (h,i \rightarrow \bot) \}. \]
The assignment $[\![ e : = 1 ] \! ]$ sets the variables $a$ and $e$ to 1;
$\mbox{Peb}(G)  \restrict [\![ e:=1 ] \! ]$ is $\mbox{Peb}(G[e:=1])$
where $G[e:=1]$ results from $G$ by removing the vertices $a$ and $e$.
If we choose the path $f \rightarrow h \rightarrow \bot$, then 
the assignment $[\![ f : = 0 ] \! ]$ sets the variables $f$ and $h$ to 0, while
the variables $a,d,e,g,i$ are all set to 1.
The set of clauses  $\mbox{Peb}(G)  \restrict [\![ f:=0 ] \! ]$ is $\mbox{Peb}(G[ f := 0 ])$, 
where $G[ f:=0 ]$ is the subgraph of $G$ containing only $b,c$ and $f$.
\end{exa}

\begin{lem}
 \label{Pebblingprooflemma}
If $G$ is a pointed graph with $n$ vertices, then $\mbox{Peb}(G)$ has a tree resolution
refutation with size $2n - 1$.
\end{lem}

\proof
Starting with the clause  $\mbox{Clause}(\bot)$ associated with the sink of $G$,
construct a sequence of purely negative clauses, working from the sink
to the sources, by successive inferences using input resolution.
Let $C \vee \neg w$ be the last clause in the sequence constructed so far, 
where $w$ is not a source vertex, and $C$ is purely negative.
Resolve $C \vee \neg w$ against the clause $\mbox{Clause}(w) = (u,v \rightarrow w)$ to
produce the next purely negative clause in the sequence, $C \vee \neg u \vee \neg v$.
The sequence must end in a purely negative clause in which all the literals are of
the form $\neg s$, where $s$ is a source vertex.
Now resolve each of these negative literals against the one-literal positive clauses
corresponding to the sources.
In this input refutation, each variable is resolved upon exactly once, so that the
refutation has size $2n - 1$.
\qed

\medskip

If $\Sigma$ is a set of clauses, then a {\em $C$-critical assignment} is a total assignment to 
the variables in $\Sigma$ that makes  all the clauses true, except $C$.
In the case of $\mbox{Peb}(G)$, we are interested in a particular family of 
critical assignments.
Let $v$ be a vertex in $G$, and $\pi = (v, \dots, t)$ a directed path in $G$ from
$v$ to the target vertex $t$.
Set all the vertices in the path $\pi$ to 0, and all other vertices in $G$ to 1.
This assignment makes all of the clauses in $\mbox{Peb}(G)$
true, except for $\mbox{Clause}(v)$.
An assignment  determined by the path $\pi$  we shall call a
{\em $v$-critical assignment}, since the clause that it falsifies
is associated with the vertex $v$.
Since we have assumed that $G$ is a pointed graph, such $v$-critical assignments
exist for all the vertices $v$ in $G$, so that $\mbox{Peb}(G)$ is minimally
inconsistent.

\begin{lem}
\label{criticalassignments}
 If $G$ is a pointed graph with $\sharp G = p$, then there are at least $p$ vertices
$v$ in $G$ for which there is a $v$-critical assignment for $\mbox{Peb}(G)$.
\end{lem}

\proof
Every pebbling strategy for $G$ must contain a configuration with $p$
pebbles, so there must be at least $p$ vertices in $G$.
For every vertex in $G$, we can construct a $v$-critical assignment
for $\mbox{Peb}(G)$ by choosing a path from $v$ to the target vertex.
\qed

\section{Constructing clause sets with large regular width}
\subsection{The basic construction}
To produce clause sets requiring large regular width, we start from 
the set of clauses $\mbox{Peb}(G)$, where $G$ is a pointed graph with
$n$ vertices.
We use the abbreviation $V$ for the set of variables $\mbox{Var}(G)$,
and $V^p$ for the set of all sequences of variables in $V$ 
of length $p$.

Let $\sigma$ be a function from $G$ to $V^p$, that is to say, a function
associating a sequence of length $p$ with every $v \in G$.
Thus, for each $v \in G$, we have an associated sequence
$\sigma(v) = \sigma_1(v), \dots, \sigma_p(v)$, where each $\sigma_j(v)$ is
a variable in $V$; the sequence may contain repetitions.
Now for $v \in G$, define the set $\mbox{Clauses}^\sigma(v)$ to be the set of 
all clauses having the form 
\[
 \mbox{Clause}(v) \vee \pm \sigma_1(v) \vee \cdots \vee \pm \sigma_p(v),
\]
where $\pm r$, for $r \in V$, is either $r$ or $\neg r$.
$\mbox{Clauses}^\sigma(v)$ contains $2^p$ 
clauses of width at most $p + 3$.
In addition, for $A \subseteq G$, define
\[
 \mbox{Clauses}^\sigma(A) = \bigcup\{\mbox{Clauses}^\sigma(v) | v \in A \}.
\]

The construction just described can be considered as an iteration of
the method used to construct the family of clauses $GT_{n,\rho}'$ defined in \S 3 of 
\cite{alekjohpiturq2007}.
A key difference from the earlier construction is that the $GT_{n,\rho}'$
examples begin from a set of clauses $GT_{n}'$ that is hard for tree resolution
(though easy for general resolution), while the present construction begins
from a set of clauses $\mbox{Peb}(G)$ that is easy for tree resolution.

The clause sets that we construct in this section are of the form
$\mbox{Clauses}^\sigma(G)$, for $G$ a pointed graph with $n$ vertices.
To ensure that these clause sets require large regular width,
the map $\sigma$ must satisfy a combinatorial condition that can 
be stated roughly as follows:
the image of any large set of vertices in $G$ has a large intersection
with any large set of variables.
In the next subsection, we  give a precise meaning to the term
``large,'' and prove the existence of a function $\sigma$ satisfying the
condition, by a probabilistic construction.

For $G$ a pointed graph, and $\sigma$ a function from $G$ to $V^p$, 
define $\mbox{Peb}^\sigma(G)$ to be $\mbox{Clauses}^\sigma(G)$.
$\mbox{Peb}^\sigma(G)$ contains 
$n \cdot 2^p$ clauses of width at most $p + 3$.
We shall show in what follows that 
the sets of clauses $\mbox{Peb}^\sigma(G)$, for an appropriate family of 
pointed graphs $G$ and functions $\sigma$, 
require large regular width, but on 
the other hand have regular tree resolution refutations
whose size is linear in $| \mbox{Peb}^\sigma(G) |$.

\subsection{A combinatorial lemma}
In this subsection, we formulate and prove the existence result described above,
by employing a probabilistic construction.
If $\sigma \in X^k$, and $B \subseteq X$, then we use the notation
$\sigma \cap B$ for the set of all elements in the sequence $\sigma$ that
also belong to the set $B$;
similarly, if $S$ is a set of such sequences, then $S \cap B$ is 
defined to be $\bigcup \{ \sigma \cap B | \sigma \in S \}$.
For $A \subseteq G$, define $\sigma(A) = \{ \sigma(v) | v \in A \}$.

\begin{lem}
 \label{existencelemma}
Let $G$ be a pointed graph with $n$ vertices, $V = \mbox{Var}(G)$
the set of variables in $\mbox{Peb}(G)$, and $p = \lceil \log^5 n \rceil$.

For any $d > 0$, and sufficiently large $n$, there is a map $\sigma$
from $G$ to $V^p$ satisfying the condition:
For all $A \subseteq G$ and $B \subseteq V$ with 
$|A| = |B| = \lfloor dn / \log n \rfloor$, 
$| \sigma(A) \cap B| \geq dn / 2 \log n$.
\end{lem}

\proof
Let us associate with each $v \in G$ a random subset of $V$ with
size $p$, chosen with replacement.
That is to say, with each $v \in G$, we associate a sequence
$\sigma(v) = \sigma_1(v), \dots, \sigma_p(v)$, 
where each variable $\sigma_j(v)$
is chosen independently and uniformly at random from the set $V$ of all variables.

In the first part of the proof, let us consider the sets $A$ and $B$
to be fixed subsets of $G$ and $V$ respectively.
Define a map $\sigma$ from $G$ to $V^p$ to be 
{\em bad for $A$ and $B$} if $| \sigma(A) \cap B| \geq dn / 2 \log n$;
otherwise {\em good for $A$ and $B$}.
We begin by proving that for fixed sets $A$ and $B$, a random
map $\sigma$ is bad with exponentially small probability.

To prove this bound on the probability, 
it is convenient to consider the construction of
the map as resulting from a series of independent choices.
Divide the sequence $1, \dots, p$ into $q = \lfloor \log^3 n \rfloor$
blocks, so that each block contains at least $\Theta ( \log^2 n )$ integers.
That is to say, the sequence $1, \dots, p$ can be written as
a concatenation $\tau_1 \tau_2, \dots, \tau_q$ of sequences
$\tau_j$, each of length at least $\Theta ( \log^2 n )$.

Fix a block $\tau_j$, where $1 \leq j \leq q$, and define a random
variable $Z$ representing the number of variables in $B$ 
that are not in the random subset $\tau_j(A)$, that is to say
\[
 Z(\tau_j) = |\{ x \in B | x \not\in  \tau_j(A) \}|.
\]
We begin by estimating the expected value of $Z$.

Let $B = \{ b_1, b_2, \dots, b_i, \dots, b_m \}$
where $m = \lfloor dn / \log n \rfloor$.
Define an indicator random variable $\Theta_i$ by:
\[
 \Theta_i (\tau_j) = 
\left\{
\begin{array}{cl}
 1, &  \mbox{if $b_i \not\in  \tau_j( A ) $} \\
0, &  \mbox {if $b_i  \in  \tau_j ( A ) $}, 
\end{array}
\right.
\]
so that $Z = \Theta_1 + \cdots + \Theta_m$.
We estimate the expected value of $\Theta_i$ by
\begin{eqnarray*}
E(\Theta_i) & \leq &   \left( 1 - \frac{1}{|V|}   \right)^{|A| \cdot |\tau_j|}\\
	  & \leq & \left(  1 - \frac{1}{ n - 1 }  \right)^{ \Theta ( n  \log n ) }  \\
	  & \leq & \exp \left(  -  \Omega \left(  \log n   \right)  \right), 
\end{eqnarray*}
showing that
\[
E(Z) \leq m \cdot \exp \left(  -   \Omega \left( \log n  \right)   \right) = m \cdot o(1) .
\]
It follows that for any given positive $\gamma$, $E(Z) < \gamma m$, for sufficiently large $n$.
For the remainder of the proof, we assume that $n$ is chosen sufficiently large so 
that $E(Z) < m/8$.

In the second stage of the proof, 
we need to show that the random variable $Z$ is
tightly concentrated around its mean.
To do this, we employ a large deviation bound for martingales, 
following \cite{kamathetal1995}.

Order the set $A$ as $\{ a_1, \dots, a_m\}$.
The sequence
$\tau_j(a_1), \tau_j(a_2), \dots, \tau_j(a_m)$
represents a random subset of variables with size 
$r = m \cdot |\tau_j| = \Theta(n \log n)$.
Let ${\mathcal R}$ be the set of all sequences in $V$ of length $r$.
For $\sigma \in {\mathcal R}$, and $1 \leq t \leq r$, define $\sigma \restrict t$ to
be the subsequence $\sigma_1, \dots, \sigma_t$.
Define an equivalence relation on $\mathcal R$ by setting,
for $\rho, \sigma \in {\mathcal R}$,
\[
  \rho \equiv_t \sigma \Longleftrightarrow \rho \restrict t = \sigma \restrict t,
\]
for $1 \leq t \leq r$, and let $\equiv_0$ be the universal relation on $\mathcal R$.
Let ${\mathcal F}_t$ be the finite Boolean algebra whose atoms are the blocks of
the partition of ${\mathcal R}$ induced by $\equiv_t$, for $0 \leq t \leq r$;
the sequence ${\mathcal F}_0, \dots, {\mathcal F}_r$ 
of Boolean algebras forms a {\em filtration} over the set $\mathcal R$.

Define a sequence of random variables $Z_0, \dots, Z_r$ by 
setting $Z_t = E( Z | {\mathcal F}_t )$.
Then $Z_0 = E(Z)$,  $Z_r = Z$, and the sequence $Z_0, \dots, Z_r$ forms
a martingale \cite[p.~221]{mcdiarmid1998}, the Doob martingale associated 
with the filtration ${\mathcal F}_0, \dots, {\mathcal F}_r$.
The intuitive picture here is that at time 0, we begin with no specific information about
a given sequence $\sigma$; we learn its values one by one at each successive time step $t$,
until we have full information about $\sigma$ at time $r$.

If $\rho$ and $\sigma$ are two sequences in $\mathcal R$ that differ at most 
at a single point, then $|Z(\rho) - Z(\sigma)| \leq 1$.
In the terminology of Alon and Spencer \cite[p.~89]{alonsp}, the random 
variable $Z$ satisfies the Lipschitz condition relative to the 
filtration ${\mathcal F}_0, \dots, {\mathcal F}_r$.
It follows by Theorem 4.1 of Chapter 7 of the monograph by Alon and Spencer \cite[p.~90]{alonsp}
that $ | Z_{t + 1} - Z_t | \leq 1 $.
Consequently, by the martingale tail inequality of Hoeffding and Azuma \cite[p.~221]{mcdiarmid1998}
\cite[p.~85]{alonsp}, 
\begin{eqnarray*}
 P( Z \geq m/2) & \leq & P( Z - E(Z) > 3m/8 ) \\
			 & < & \exp( - (3m/8)^2 / 2r ) \\
			 & \leq  &    \exp( - \Omega  ( n / \log^3 n  ) ).
\end{eqnarray*}

Let $W$ be the random variable representing the number of variables in
$B$ not in the image of $A$ under $\sigma$:
\[
	W(\sigma) = | \{  x \in B | \: x \not\in  \sigma( A )    \:  \} |.
\]
Since the maps $\tau_1, \dots, \tau_q$ are constructed independently, it follows that
\[
 P( W \geq m/2) \leq   [ \exp( - \Omega  ( n / \log^3 n  ) ) ]^q =   \exp( - \Omega  ( n  ) ).
\]

We can now complete the proof of the existence of a map $\sigma$ satisfying the condition of
Lemma \ref{existencelemma}.
The probability that a random map $\rho \in { \mathcal R }$ is bad for some $A$ and $B$ is bounded
by 
\[
\binom{n}{ m }^2  \exp( - \Omega  ( n  ) ).
\]
Using the simple inequality 
\[
\binom{n}{k} \leq \left(  \dfrac{en}{k} \right)^k, 
\]
found in Bollob\'{a}s's textbook on graph theory 
\cite[p.~216]{bollobas1998}, the binomial coefficient above can be bounded by
\begin{eqnarray*}
 \binom{n}{ m } &  \leq  & 
		\left(  \dfrac{en}{m} \right)^{m} \\
	& = & \left( e^{ O( \log \log n) } \right)^{O(n/ \log n)} \\
	& = & e^{ O ( n \log \log n / \log n )}.
\end{eqnarray*}
Hence, the probability can be bounded above by
\[
  \exp(  O ( n  \log \log n / \log n)  )  \exp( - \Omega  ( n  ) ) =  \exp( - \Omega  ( n  )).
\]
Consequently, the probability that a random map $\rho$ is bad for some $A$ and $B$ is 
exponentially small for sufficiently large $n$, showing that a map satisfying the condition of 
the lemma must exist.
\qed

\section{Separating regular size and width}
\label{separationsection}

Let $G$ be a pointed graph,
$V$ the set of vertices in $G$ (other than the sink) and $N$ and $p$ positive integers.
We define a map $\sigma$ to be {\em good for $G$, $N$, and $p$}
if it satisfies the condition:
There is a map $\sigma$ from $G$ into $V^p$ so that for any $A \subseteq G$ and $B \subseteq V$,
if $|A| =|B| = \lfloor N \rfloor$, then $|\sigma(A) \cap B| \geq N/2$.
Lemma \ref{existencelemma} states that for any $d > 0$, given sufficiently large $n$, 
$p = \lceil \log^5 n \rceil$, and $N = d n / \log n$, for every pointed graph with $n$ vertices,
there is a map $\sigma$ that is good for $G$, $N$, and $p$.

This lemma allows to construct a set of examples that have polynomial-size regular resolution
refutations, but large regular width.
The construction is based on the following result of Paul, Tarjan and Celoni.
\begin{thm}
\label{pct}
\cite{paulcelonitarjan1977}
There is a sequence of binary pointed graphs $G_1, \dots, G_i, \dots$ 
with pebbling number at least $c n(i) / \log n(i) $, for sufficiently large $i$, where
$n(i) =| G_i | =  O(i 2^i)$, and $c > 1/20$.
\end{thm}

It should be mentioned that the graphs $H_i$ constructed by Paul, Tarjan and Celoni,
though binary, are not pointed, since they are constructed to have multiple sink nodes.
However, in their main theorem, they show that for sufficiently large $i$,
their graph $H_i$ contains a sink node that requires $c n(i) / \log n(i) $ pebbles to pebble
it, starting from the empty configuration.
Hence, we can construct a pointed graph from $H_i$ by choosing such a sink node,
and considering the subgraph $G_i$ containing all the nodes from which this sink
is accessible.
This subgraph $G_i$ still satisfies the condition $n(i) =| G_i | =  O(i 2^i)$,
so the main theorem of Paul, Tarjan and Celoni continues to hold, if we add 
the qualifier ``pointed.''

\begin{lem}
\label{upperboundlemma}
Let $G$ be a pointed graph with $n$ vertices, and $\sigma$ a map from $G$ to $V^p$, where $p = \lceil \log^5 n \rceil$.
Then the set of clauses $\mbox{Peb}^\sigma(G)$ contains $n-1$ variables and $n^{O(\log^4 n)}$ clauses,
and has a regular tree refutation with size $n^{O(\log^4 n)}$.
\end{lem}

\proof
By Lemma \ref{Pebblingprooflemma}, $\mbox{Peb}(G)$ has an input refutation 
with size at most $2n - 1$.  
For a given vertex $v$ in $G$, the clause $C(v)$ associated with the vertex
can be derived from $\mbox{Clauses}^\sigma(v)$ by a tree resolution proof
with size $2^{O(\log^5 n)} = n^{O(\log^4 n)}$.
Consequently, $\mbox{Peb}^\sigma(G_i)$ has a tree refutation with size 
$O(n) \cdot n^{O(\log^4 n)}$, that is, $n^{O(\log^4 n)}$.
This tree refutation may not be regular; however, if irregularities are present,
it is possible to remove them \cite{ts70} \cite[p.\ 436]{urq95} 
resulting in a smaller regular tree-style refutation.
\qed

\begin{lem}
\label{lowerboundlemma}
Let $G$ be a pointed graph with pebbling number $\sharp G = N$ and $\sigma$ a map 
that is good for $G$, $N/2$, and $p$.
Then any regular resolution refutation of $\mbox{Peb}^\sigma(G)$ must contain a clause with width
at least $N/4$.
\end{lem}

\proof
We prove the Lemma by showing that 
the Adversary wins the regular $N/4$-width game based on $\mbox{Peb}^\sigma(G)$.
The winning strategy has two stages.
In the first stage, the Adversary maintains a pebbling assignment to $G$;
at the start of the game, this assignment is empty.
In the second stage, the Adversary answers according to a fixed 
$v$-critical assignment.

Assume that it is the Adversary's turn, that $\pi$ is the current pebbling  
assignment to the variables $V$ in $\mbox{Peb}^\sigma(G)$, and that $x$ is
the variable currently queried by the Prover.
The Adversary answers the current query according to these rules.
\begin{enumerate}[(1)]
 \item If the variable $x$ is already assigned a value by $\pi$, then
answer the query according to $\pi$;
\item 
If the variable queried is not assigned a value by $\pi$, then it
must be associated with a node $v \in G \restrict \pi$.
Extend $\pi$ to a pebbling assignment $\pi'$ so that
$\pi'$ contains  $[\![ v : = b ] \! ]$, choosing $b$ so as to maximize
the pebbling number of $G \restrict \pi'$.
\end{enumerate}
The Adversary continues to play according to these rules until $\lfloor N/2 \rfloor$
nodes in $G$ have been queried; when this happens, the first stage is completed.

With the first stage completed, let $\alpha$ be the current extended assignment,
$\pi$ the current pebbling assignment maintained by the Adversary;
we assume that it is the Prover's turn.
By Lemma \ref{criticalassignments}, there are at least $N/2$ vertices
$v \in G \restrict \pi$ for which there is a $v$-critical assignment
for $\mbox{Peb}( G \restrict \pi)$.
If $\phi$ is such a critical assignment, then $\pi \cup \phi$ is
a $v$-critical assignment for $\mbox{Peb}(G)$.
Let $A$ be the set of all nodes in $G$ satisfying this condition, and
$B$ the set of variables queried in the game so far.
Because $|A|, |B| \geq \lfloor N/2 \rfloor$,
$ | \sigma(A) \cap B | \geq  N/4$, since $\sigma$
is good for $G$, $N$, and $p$.

Since the Prover and Adversary are playing the regular $N/4$-width game,
it follows that $|\alpha| < N/4$ (since the current assignment after
the Adversary's reply has width $|\alpha| + 1$).
Hence, at least one variable $v$ in $\sigma(A) \cap B$ must be forgotten
in $\alpha$.
Let $\phi$ be a $v$-critical assignment for $\mbox{Peb}(G)$;
$\phi$ is also a $v$-critical assignment for $\mbox{Peb}^\sigma(G)$.
In the second stage of the strategy, the Adversary answers all queries in
accordance with the assignment $\phi$.
Since $\phi$ makes all of the clauses in $\mbox{Peb}^\sigma(G)$ true,
except for a clause in $\mbox{Clauses}(v)$ containing the variable $v$,
this strategy results in a win for the Adversary, since the variable
$v$ is forgotten, so the Prover cannot query it again.
\qed

\medskip

\begin{thm}
\label{maintheorem}
 There is an infinite sequence $\Sigma_1, \Sigma_2, \dots, \Sigma_i, \dots$ of contradictory sets of clauses
and a corresponding list of parameters $n(1), n(2), \dots, n(i), \dots$ so that (abbreviating $n(i)$ as $n$):
\begin{enumerate}[\em(1)]
 \item Each clause set $\Sigma_i$ contains $n -1$ variables and  $n^{O(\log^4 n)}$ clauses
with width $O( \log^5 n )$;
\item $\Sigma_i$ has a regular tree refutation with size $n^{O(\log^4 n)}$;
\item Any regular refutation of $\Sigma_i$ must contain a clause with width $\Omega ( n / \log n) $.
\end{enumerate}
\end{thm}

\proof
Define $\Sigma_i = \mbox{Peb}^\sigma(G_i)$, where $G_i$ is one
of the sequence of pointed graphs in Theorem \ref{pct}.
The theorem follows by Lemmas \ref{existencelemma}, \ref{upperboundlemma} and \ref{lowerboundlemma}. 
\qed

\medskip

Although the clause sets in Theorem \ref{maintheorem} have size quasi-polynomial in $n$,
they have regular tree refutations that are linear in the size of the clause sets themselves.
Furthermore, if we compute the significant quantities in the second part of Theorem \ref{BenSassonWigdersonTheorem},
we find that if the corresponding theorem held for regular size and width, then
regular refutations of these clause sets would have to have size exponential in $n / \log^2 n$.
This shows that the relations between size and width holding for tree resolution and general resolution
cannot be generalized to the case of regular resolution.

\section*{Acknowledgments}

I wrote this paper for a five day workshop on proof complexity at the Banff International 
Research Station in October 2011.
I would like to express my thanks to the organizers, Sam Buss, Stephen Cook, Antonina 
Kolokolova, Toni Pitassi and Pavel Pudl\'{a}k for a most stimulating workshop, and also
to Paul Beame, who, following my talk, pointed out a computational error in the original version
of the paper.

\end{document}